\documentclass[lettersize,journal]{IEEEtran}
\usepackage{amsmath,amsfonts}
\usepackage{algorithmic}
\usepackage{algorithm}
\usepackage{array}
\usepackage{bm}
\usepackage{booktabs}   
\usepackage{amsmath,amssymb} 
\newcommand{\R}{\mathbb{R}}
\usepackage[caption=false,font=footnotesize]{subfig}

\usepackage{textcomp}
\usepackage{xcolor}
\usepackage{stfloats}
\usepackage{url}
\usepackage{verbatim}
\usepackage{graphicx}
\usepackage{cite}
\usepackage{titlesec}

\titlespacing{\subsection}{0pt}{\parskip}{-\parskip}
\begin{document}
\title{Sensing-Assisted Adaptive Beam Probing with Calibrated Multimodal Priors and Uncertainty-Aware Scheduling}

\author{Abidemi Orimogunje, Vukan Ninkovic, Ognjen Kundacina, Hyunwoo Park, Sunwoo Kim, Dejan Vukobratovic, Evariste Twahirwa, and Gaspard Gashema
\thanks{This paper was produced by the IEEE Publication Technology Group. They are in Piscataway, NJ.}
\thanks{Manuscript received April 19, 2021; revised August 16, 2021.}}

\markboth{Journal of \LaTeX\ Class Files,~Vol.~14, No.~8, August~2021}%
{Shell \MakeLowercase{\textit{et al.}}: A Sample Article Using IEEEtran.cls for IEEE Journals}

\IEEEpubid{0000--0000/00\$00.00~\copyright~2021 IEEE}

\maketitle

\begin{abstract}
Highly directional mmWave/THz links require rapid beam alignment, yet exhaustive codebook sweeps incur prohibitive training overhead. This letter proposes a sensing-assisted adaptive probing policy that maps multimodal sensing (radar/LiDAR/camera) to a calibrated prior over beams, predicts per-beam reward with a deep Q-ensemble whose disagreement serves as a practical epistemic-uncertainty proxy, and schedules a small probe set using a Prior-Q upper-confidence score. The probing budget is adapted from prior entropy, explicitly coupling sensing confidence to communication overhead, while a margin-based safety rule prevents low signal-to-noise ratio (SNR) locks. Experiments on DeepSense-6G (train: scenarios 42 and 44; test: 43) with a 21-beam discrete Fourier transform (DFT) codebook achieve Top-1/Top-3 of 0.81/0.99 with $\mathbb{E}[K]\approx2$ probes per sweep and zero observed outages at $\theta=0$~dB with $\Delta=3$~dB. The results show that multimodal priors with ensemble uncertainty match link quality and improve reliability compared to ablations while cutting overhead with better predictive model. 
\end{abstract}
\section{Introduction}
\IEEEPARstart{D}{irectional} millimeter wave (mmWave) and terahertz (THz) communication systems rely on narrow beams to overcome severe path loss, but aligning these beams during initial access or beam tracking incurs large training overhead~\cite{8458146, liu2021beam, 9869437, 9652539, 10720905}. In conventional frameworks, the transmitter and receiver must sweep through different beam directions to find the best alignment, leading to latency, energy expenditure, and potential link outage if alignment fails~\cite{sun2015beam,7880676, 9131853}. To handle these challenges, future 6G systems are expected to integrate communication and sensing functionalities, leveraging external sensors to aid beam alignment~\cite{10422712, charan2022vision}. 

Recent research has shown that multimodal sensing features from camera, radar, light detection and ranging (LiDAR), and global positioning system (GPS) sensors can be exploited for beam alignment and to predict the optimal beam, thereby minimizing the need for exhaustive search \cite{10735366, 10719654, zheng2025m2beamllmmultimodalsensingempoweredmmwave}. {For example, using camera images and position data to guide beam selection yields $75\,\%$ of correct beam ranked first (top-1 accuracy) and $100\,\%$ correct beam included within the top three predictions i.e. top-3 accuracy in realistic vehicular scenarios~\cite{charan2022vision}. Likewise, radar-aided beam prediction has demonstrated over 90\,\% top-5 prediction accuracy (correct beam prediction within top 5 predictions) while saving $93\,\%$ of the beam training overhead~\cite{demirhan2022radar}. 

These sensor-aided approaches highlight the potential of using external multi-modal sensing sources to drastically cut alignment overhead and latency \cite{10561505, 9765510}. However, previous works often optimizes beam classification accuracy but (i) does not explicitly quantify uncertainty to control probing resources, and (ii) lacks a reliability mechanism that reports outage-related metrics under a clear policy.
In this letter, we propose a sensing-assisted beam-probing policy that couples sensing confidence to communication probing overhead with the following contributions:
\begin{itemize}
\item A calibrated multimodal beam prior that converts sensing features into a per-sweep probability mass function (PMF) over beams (with temperature scaling driven by prior sharpness).
\item A Q-ensemble that predicts per-beam reward and uses ensemble disagreement as an epistemic uncertainty proxy~\cite{lakshminarayanan2017simplescalablepredictiveuncertainty}.
\item A upper confidence bound (UCB)-style scheduler that selects a small probing set $\mathcal{S}_t$ and adapts $K_t$ from the entropy of the calibrated prior, making the sensing--communication resource coupling explicit.
\item A lightweight safety rule and reporting protocol that logs both threshold-outage and margin-outage, plus the activation frequency of the safety mechanism.
\end{itemize}
\section{System Model and Problem Formulation}
We consider a single downlink link where an access point (AP) performs beam selection from a discrete Fourier transform (DFT) codebook of size $B$, with beam index set $\mathcal{B}=\{1,\ldots,B\}$.
At each sweep $t$, an AP-based platform provides \emph{sensing side information} in the form of a fused feature vector $\bm{x}_t$ (from frequency modulated continuous wave (FMCW) radar, camera, and LiDAR processing).
The AP can probe only a \emph{small} subset of beams $\mathcal{S}_t\subset\mathcal{B}$ with cardinality $|\mathcal{S}_t|=K_t\ll B$ by transmitting pilots and collecting per-beam in-phase and quadrature (IQ) recordings $\bm{Z}_{t,b}$ for $b\in\mathcal{S}_t$.
Here, $K_t$ is the \emph{beam-training overhead} (control-plane cost) per sweep.
We treat sensing as given side information (i.e., sensing-assisted communications) and focus on how sensing confidence can reduce communication probing overhead while preserving link reliability.
\subsection{SNR proxy and oracle beam}
From the IQ recording $\bm{Z}_{t,b}$, we compute a robust absolute-SNR proxy using a percentile power ratio:
\begin{equation}
\mathrm{SNR}_{\mathrm{dB}}(\bm{Z}_{t,b})=10\log_{10}
\left(
\frac{\mathrm{perc}_{p_s}\!\bigl(|\bm{Z}_{t,b}|^2\bigr)}
{\mathrm{perc}_{p_n}\!\bigl(|\bm{Z}_{t,b}|^2\bigr)+\varepsilon}
\right),
\label{eq:snrproxy}
\end{equation}
with $(p_s,p_n)=(99.7,20)$ and $\varepsilon>0$ for numerical stability.
For reporting Top-$k$ metrics, we define the \emph{oracle} best beam as the maximizer over the full codebook:
\begin{equation}
b_t^\star=\arg\max_{b\in\mathcal{B}}~\mathrm{SNR}_{\mathrm{dB}}(\bm{Z}_{t,b}).
\label{eq:oracle_beam}
\end{equation}
Importantly, $b_t^\star$ is used only to form evaluation/training targets (available in the dataset via exhaustive logging), whereas the online policy probes only $K_t$ beams.

\subsection{Probing decision, locking rule, and reliability metrics}
Given a probed set $\mathcal{S}_t$, the best probed candidate is
\begin{equation}
b_t^{\mathrm{best}}=\arg\max_{b\in\mathcal{S}_t}~\mathrm{SNR}_{\mathrm{dB}}(\bm{Z}_{t,b}).
\label{eq:b_best}
\end{equation}
To avoid low-margin locks, we apply a lightweight safety shield with outage threshold $\theta$ and safety margin $\Delta\ge 0$:
\begin{equation}
b_t^{\mathrm{lock}} =
\begin{cases}
  b_t^{\mathrm{best}}, & \text{if } \mathrm{SNR}_{\mathrm{dB}}(\bm{Z}_{t,b_t^{\mathrm{best}}}) \ge \theta + \Delta,\\[3pt]
  b_t^{\mathrm{safe}}, & \text{otherwise},
\end{cases}
\label{eq:b_lock}
\end{equation}
where $b_t^{\mathrm{safe}}$ is selected from a local neighbor set around the previous lock to preserve continuity.
Let $\mathcal{N}(b_{t-1}^{\mathrm{lock}})=\{b'\in\mathcal{B}:\mathrm{circ\_dist}(b',b_{t-1}^{\mathrm{lock}})\le w\}$ denote a circular neighborhood of width $w$; the shield picks the neighbor with highest \emph{last-known} $\mathrm{SNR}_{\mathrm{dB}}$ among those satisfying $\mathrm{SNR}_{\mathrm{dB}}\ge \theta+\Delta$ (else it retains $b_{t-1}^{\mathrm{lock}}$).

To transparently quantify reliance on the safety fallback, we define the shield activation indicator
$\mathbb{I}_t^{\mathrm{shield}}=\mathbf{1}\{b_t^{\mathrm{lock}}\neq b_t^{\mathrm{best}}\}$,
and report the \emph{shield activation rate} $\mathbb{E}[\mathbb{I}_t^{\mathrm{shield}}]$ in the results.
We also define the threshold-outage event for a locked beam as
\begin{equation}
\mathbb{I}_t^{\mathrm{out}}=\mathbf{1}\{\mathrm{SNR}_{\mathrm{dB}}(\bm{Z}_{t,b_t^{\mathrm{lock}}})<\theta\},
\label{eq:outage_theta}
\end{equation}
and report its empirical rate to quantify reliability under the chosen $\theta$.

For offline learning, we form a per-sweep IQ-derived reward vector $\bm{R}^{\mathrm{IQ}}_t\in\mathbb{R}^{B}$ with $[\bm{R}^{\mathrm{IQ}}_t]_b=\mathrm{SNR}_{\mathrm{dB}}(\bm{Z}_{t,b})$.
We optionally reduce label variance by blending IQ rewards with a sensing-derived beam prior (before calibration) to form a normalized hybrid target $\bm{R}_t^{\mathrm{hyb}}$ used to train the Q-ensemble:
\begin{equation}
\bm{R}^{\mathrm{hyb}}_{t}
=
\alpha\,\mathrm{zscore}_t(\bm{R}^{\mathrm{IQ}}_t)
+(1-\alpha)\,\mathrm{zscore}_t(\bm{\pi}^{\mathrm{sense}}_t),
\label{eq:hybrid}
\end{equation}
where $\alpha\in[0,1]$, $\mathrm{zscore}_t(\cdot)$ denotes per-sweep standardization across beams and $\bm{\pi}^{\mathrm{sense}}_t$ is a beam prior derived from sensing.
\subsection{Problem formulation: overhead-reliability and link-quality trade-off}
Given sensing features $\bm{x}_t$ (and past decisions/measurements), the policy $\pi^{\mathrm{sel}}$ selects a probing set $\mathcal{S}_t$ (equivalently $K_t$ and which beams to probe) and outputs the final locked beam $b_t^{\mathrm{lock}}$.
Our objective is communication-centric: maximize locked-beam quality while controlling beam-training overhead and outage risk.
This is stated as the constrained problem and equivalently, a penalized form
\begin{align}
\textbf{(P1) Constrained:}\quad
\max_{\pi^{\mathrm{sel}}}~ \mathbb{E}\!\left[\mathrm{SNR}_{\mathrm{dB}}(\bm{Z}_{t,b_t^{\mathrm{lock}}})\right]
& \notag \\
\text{s.t.}~ \mathbb{E}[K_t]\le\bar{K},\ 
\Pr\left\{ \mathrm{SNR}_{\mathrm{dB}}(\bm{Z}_{t,b_t^{\mathrm{lock}}})<\theta \right\}\le\delta
&,
\end{align}
\begin{align}
\textbf{(P2) Penalized:}\quad
\max_{\pi^{\mathrm{sel}}}~ \mathbb{E}\Big[\, \mathrm{SNR}_{\mathrm{dB}}(\bm{Z}_{t,b_t^{\mathrm{lock}}})\,
& \notag \\
-~ c_K K_t 
-~ c_{\mathrm{out}}\mathbf{1}\left\{ \mathrm{SNR}_{\mathrm{dB}}(\bm{Z}_{t,b_t^{\mathrm{lock}}})<\theta \right\} \Big] &.
\end{align}
where $\bar{K}$ is an average probing-budget constraint and $\delta$ is an outage-risk constraint.
This formulation makes the sensing--communication coupling explicit: sensing affects the policy's confidence about beams (via a sensing-derived prior), and that confidence is used to adapt $K_t$ so as to satisfy the overhead constraint while maintaining reliability.
\section{Proposed Method}\label{sec:methods}
\subsection{Calibrated multimodal prior}
A multimodal prior network maps sensing features $\bm{x}_t$ to logits $z_t(b)\in\R$ for $b\in\mathcal{B}$.
We standardize logits using training statistics $(\mu_{\mathrm{train}},\sigma_{\mathrm{train}})$:
\begin{equation}
\tilde z_t(b)=\frac{z_t(b)-\mu_{\mathrm{train}}}{\sigma_{\mathrm{train}}+\epsilon}.
\label{eq:logit_standardize}
\end{equation}
Temperature scaling \cite{guo2017calibration} yields a calibrated PMF:
\begin{equation}
p_t(b)=\frac{\exp(\tilde z_t(b)/T_{\mathrm{eff}})}{\sum_{b'\in\mathcal{B}}\exp(\tilde z_t(b')/T_{\mathrm{eff}})}.
\label{eq:softmax}
\end{equation}
where effective temprature,
$T_{\mathrm{eff}}
= \mathrm{clip}\left(
T_0 \left(\frac{\hat{s}_t}{s_{\min}}\right)^{-\alpha} \big(1 + \gamma\,u^{\mathrm{norm}}_t\big),
\,T_{\min},\,T_{\max}\right)$. Also, we compute the prior entropy
\begin{equation}
H(p_t)=-\sum_{b\in\mathcal{B}}p_t(b)\log p_t(b),
\label{eq:entropy}
\end{equation}
and use it to set the probing budget $K_t$ (Sec.~\ref{subsec:scheduling}).
\subsection{Q-ensemble reward prediction and uncertainty}
We use an ensemble of $M$ predictors $\{Q_m\}_{m=1}^M$, where $Q_m(\bm{x}_t,b)$ predicts a scalar reward proxy.
Ensemble mean and standard deviation are follows:
\begin{align}
\mu_t(b) &= \frac{1}{M}\sum_{m=1}^M Q_m(\bm{x}_t,b), \label{eq:q_mean}\\[2pt]
\tau_t(b) &= \sqrt{\frac{1}{M}\sum_{m=1}^M\big(Q_m(\bm{x}_t,b)-\mu_t(b)\big)^2}. \label{eq:q_std}
\end{align}
The $\tau_t(b)$ is used as a practical epistemic uncertainty proxy in the standard deep-ensemble sense~\cite{lakshminarayanan2017simplescalablepredictiveuncertainty}; we do not claim formal regret guarantees under general non-stationary mmWave dynamics, but we validate its utility empirically through ablations (Sec.~V).
For scheduling we normalize per sweep uncertainty:
\begin{equation}
\widehat{\sigma}_t(b)=\frac{\tau_t(b)}{\max_{b'\in\mathcal{B}}\tau_t(b')+\epsilon}.
\label{eq:unc_norm}
\end{equation}
\subsection{Prior-Q UCB scheduling and adaptive probing}\label{subsec:scheduling}
We score each beam by
\begin{equation}
s_t(b)=(1-\lambda)\,\mathrm{zscore}_t(\mu_t(b))+\lambda\log p_t(b)+\beta\,\widehat{\sigma}_t(b),
\label{eq:score}
\end{equation}
where $\lambda\in[0,1]$ and $\beta\ge0$.
We form $\mathcal{S}_t$ by ranking beams in descending $s_t(b)$ and greedily selecting beams while enforcing a minimum circular separation $d_\theta$.

\subsection{Adaptive probe budget and Safety shield}
We set $K_t$ deterministically from the prior entropy:
\begin{equation}
K_t=
\begin{cases}
K_{\min}, & H(p_t)\le H_{\mathrm{low}},\\
K_{\max}, & H(p_t)\ge H_{\mathrm{high}},\\
K_{\mathrm{mid}}, & \text{otherwise},
\end{cases}
\label{eq:K_rule}
\end{equation}
with $K_{\min}<K_{\mathrm{mid}}<K_{\max}$ and $H_{\mathrm{low}}<H_{\mathrm{high}}$.

While the entropy $H(p_t)$ captures global prior uncertainty, it may not detect near-ties among the most likely beams.
We therefore sort the top prior masses $p_t^{(1)}\ge p_t^{(2)}\ge p_t^{(3)}$ and define
$g_t \triangleq p_t^{(1)}-p_t^{(3)}$.
If $g_t<\tau_{\mathrm{gap}}$, we increase the probing budget by one beam,
$K_t \leftarrow \min\{K_t+1,\,K_{\max}\}$.
Setting $\tau_{\mathrm{gap}}=0$ disables this step.
The complete per-sweep procedure (prior calibration, scoring, adaptive $K_t$, diverse set selection, probing, and safety lock) is summarized in Algorithm~\ref{alg:prior_qucb}.
We use $\theta=0$\,dB as a conservative connectivity boundary (signal and noise powers comparable) and $\Delta=3$\,dB as a standard engineering safety margin (approximately a factor-of-two power buffer) to avoid low-margin locks.
We report both outage rate \eqref{eq:outage_theta} and shield activation rate to quantify how often reliability is provided by fallback.
\subsection{Computational complexity}
Per sweep, inference requires one prior forward pass and $M$ ensemble forward passes producing $B$ scores, i.e., $\mathcal{O}(MB)$, plus sorting $\mathcal{O}(B\log B)$.
The dominant practical cost in beam training is measurement overhead, which scales with only $K_t\ll B$ probes (versus exhaustive sweeping $K=B$).
\begin{algorithm}[t]
\small
\caption{Sensing-assisted Prior--Q UCB scheduling with adaptive probing and safety shield (per sweep $t$)}
\label{alg:prior_qucb}
\begin{algorithmic}[1]
\STATE \textbf{Inputs:} fused feature $\bm{x}_t$; prior logits $z_t(\cdot)$; Q-ensemble $\{Q_m\}_{m=1}^M$; codebook $\mathcal{B}$; previous locked beam $b_{t-1}^{\mathrm{lock}}$.
\STATE \textbf{Parameters:} $\lambda,\beta$; $d_\theta$; $H_{\mathrm{low}},H_{\mathrm{high}}$;$K_{\min},K_{\mathrm{mid}},K_{\max}$; $\tau_{\mathrm{gap}}$; $\theta,\Delta$; neighbor width $w$; $\epsilon$.
\STATE \textbf{(A) Prior:} compute $\tilde z_t(b)$ via \eqref{eq:logit_standardize}; compute $p_t(b)$ via \eqref{eq:softmax}; compute $H(p_t)$ via \eqref{eq:entropy}.
\STATE \textbf{(B) Q-ensemble:} compute $\mu_t(b),\tau_t(b)$ via \eqref{eq:q_mean}--\eqref{eq:q_std}; normalize $\widehat{\sigma}_t(b)$ via \eqref{eq:unc_norm}.
\STATE \textbf{(C) Score:} compute $s_t(b)$ via \eqref{eq:score}.
\STATE \textbf{(D) Budget:} set $K_t$ via \eqref{eq:K_rule}; optional bump if $p_t^{(1)}-p_t^{(3)}<\tau_{\mathrm{gap}}$.
\STATE \textbf{(E) Select:} build $\mathcal{S}_t$ by descending $s_t(b)$ with min separation $d_\theta$ until $|\mathcal{S}_t|=K_t$.
\STATE \textbf{(F) Probe:} measure $\mathrm{SNR}_{\mathrm{dB}}(\bm{Z}_{t,b})$ via \eqref{eq:snrproxy} for $b\in\mathcal{S}_t$; set $b_t^{\mathrm{best}}$ via \eqref{eq:b_best}.
\STATE \textbf{(G) Lock:} lock $b_t^{\mathrm{lock}}$ via \eqref{eq:b_lock}.
\STATE \textbf{Output:} $b_t^{\mathrm{lock}}$, $\mathcal{S}_t$, $K_t$.
\STATE \textbf{Complexity (per sweep):} compute $\mathcal{O}(MB+B\log B)$; shield scan $\mathcal{O}(w)$; probing cost is $K_t$ beam measurements with $K_t\ll B$.
\end{algorithmic}
\end{algorithm}
\section{Experimental Setup and Evaluation Protocol}
We evaluate on the real‑world DeepSense‑6G dataset~\cite{alkhateeb2023deepsense}. All models were run with CPU inference for fair comparison of complexity. The Top-1 and Top-3 accuracies are computed with respect to the DFT codebook (oracle), while our online policy probes only $K_t \ll B$ beams per sweep. We focus on four key performance metrics: Top-1/Top-3 locked-beam accuracy, average probes per sweep, outage probability, and the mean with 5th-percentile ($p05$) of the locked-beam SNR.
\subsubsection{Dataset and Split} We use DeepSense‑6G scenarios 42 and 44 for training, while scenario 43 is used for testing. Each sweep uses a $B=21$ beam codebook. The state vector $\bm{x}_t$ fuses radar features (64\(\times\)64 heatmap compression), LiDAR polar histogram and bearing, a camera 21‑bin horizontal edge histogram and bearing, and meta features. Training statistics (feature standardization and prior logit statistics) are saved and reused at evaluation.
\subsubsection{Training and Inference} The prior is optimized with class‑weighted cross‑entropy and soft targets derived from the per‑sweep IQ‑SNR rewards; standardized feature scaling is persisted and reused at test time. Temperature scaling parameters are learned from global prior‑logit statistics saved during training. 
\subsubsection{Evaluation Metrics}
The following metrics are define to evaluate the performance of our framework:\\
(i) Top‑1/Top‑3: the fraction of sweeps in which $b_t^{\mathrm{lock}}$ equals the oracle best beam (Top‑1) or lies among the top‑3 beams by oracle SNR (Top‑3).\\
(ii) Overhead: mean probes per sweep, $\mathbb{E}[K_t]$.\\
(iii) Outage: $\Pr\{\mathrm{SNR}_{\mathrm{dB}}(\bm{Z}_{t,b_t^{\mathrm{lock}}})<\theta\}$ with $\theta{=}0$\,dB.\\
(iv) SNR statistics: mean and p05 of the locked‑beam absolute SNR proxy.
\section{Results and Discussion}\label{sec:results}
\subsection{Main performance and baselines}
Table~\ref{tab:main_perf} reports locked-beam Top-1/Top-3 accuracy, average probing overhead $\mathbb{E}[K_t]$, and locked-beam SNR/outage under the probe-and-lock protocol defined in Sec.~II.
We compare against Random selection and a prior-only argmax baseline (selecting $\arg\max_b p_t(b)$ with $K_t{=}1$), and a contextual bandit baseline (LinUCB) at a matched probing budget ($K_t{=}2$).
To isolate the contribution of adaptive budgeting, we additionally include a fixed-budget control for our method ($K_t\equiv2$).

The proposed sensing-assisted Prior--Q policy achieves Top-1/Top-3 $=0.81/0.99$ with $\mathbb{E}[K_t]=1.99$, corresponding to a $B/\mathbb{E}[K_t]\approx 21/1.99 \approx 10.6\times$ reduction in probing relative to exhaustive sweeping ($K_t=B=21$), while maintaining $0\%$ outage at $\theta=0$\,dB with safety margin $\Delta=3$\,dB.
Fig.~\ref{fig:snr_cdf} shows that the locked-beam SNR proxy in~\eqref{eq:snrproxy} is tightly concentrated (mean $14.18$\,dB; p05 $14.13$\,dB), consistent with stable link quality under very low measurement overhead.
The fixed-$K_t$ control matches the adaptive result, indicating that performance is driven primarily by the learned beam ranking (calibrated prior fused with Q-ensemble prediction and an uncertainty bonus), while the entropy rule mainly selects a low-overhead operating point.
\begin{table}[t]
\vspace{-1em} 
\centering
\caption{Main performance on DeepSense-6G scenario~43 (oracle defined over $B=21$ beams) under the probe-and-lock protocol}
\label{tab:main_perf}
\setlength{\tabcolsep}{2pt}
\begin{tabular}{lcccccc}
\toprule
Method & Top-1 & Top-3 & $\mathbb{E}[K_t]$ & SNR$_{\rm avg}$ & SNR$_{\rm p05}$ & Outage \\
\midrule
Random ($K_t{=}1$) & 0.03 & 0.14 & 1.00 & -- & -- & -- \\
Prior-only argmax ($K_t{=}1$) & 0.56 & 0.83 & 1.00 & -- & -- & -- \\
LinUCB ($K_t{=}2$) & 0.08 & 0.27 & 2.00 & 14.17 & 14.13 & 0.00 \\
Ours (fixed $K_t{=}2$) & 0.81 & 0.99 & 2.00 & 14.18 & 14.13 & 0.00 \\
\textbf{Ours (adaptive $K_t$)} & \textbf{0.81} & \textbf{0.99} & \textbf{1.99} & \textbf{14.18} & \textbf{14.13} & \textbf{0.00} \\
\bottomrule
\end{tabular}
\vspace{-1em} 
\end{table}
\begin{figure}[t]
    \centering
    \includegraphics[width=0.8\linewidth]{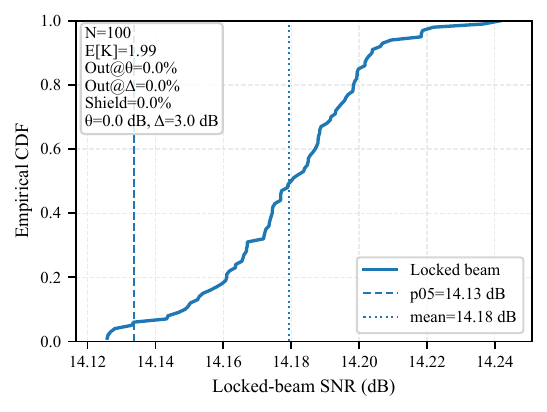}
    \caption{Empirical CDF of the locked-beam SNR proxy \eqref{eq:snrproxy} for the proposed adaptive policy on DeepSense-6G scenario~43 ($B{=}21$). Vertical markers denote the mean and the 5th-percentile (p05), illustrating a tight locked-beam SNR distribution under $\mathbb{E}[K_t]\!\approx\!2$ probes per sweep.}
    \label{fig:snr_cdf}
\end{figure}
\subsection{Ablation: score terms and safety mechanism}
Fig.~\ref{fig:ablations} isolates the contribution of the probing-and-lock components at a matched budget ($K_t{=}2$).
The full policy (prior--Q fusion with uncertainty bonus and the margin-based lock rule) maintains high Top-1/Top-3 accuracy while preserving the same high-SNR operating point.
Importantly, the measured shield activation rate is negligible in this scenario (consistent with Fig.~\ref{fig:snr_cdf}), indicating that the reported reliability is not achieved by frequent rule-based overrides; rather, the learned ranking and targeted probing already produce high-margin candidates}.
\begin{figure}[t]
    \centering
    \includegraphics[width=1.06\linewidth]{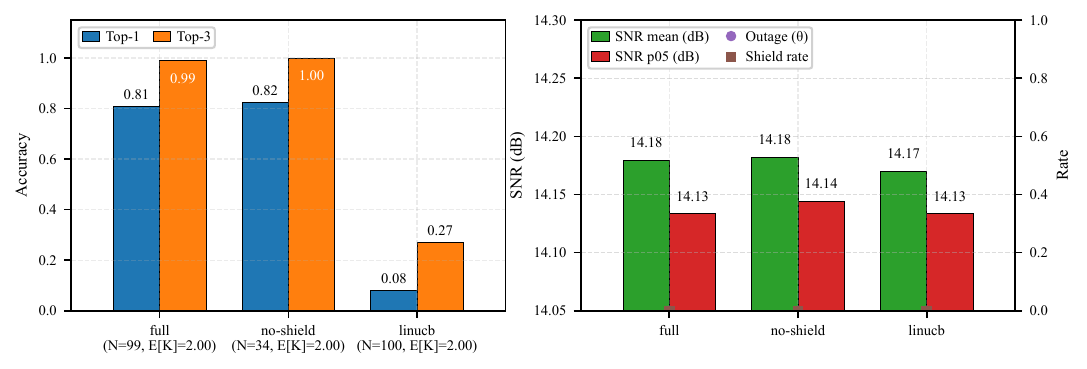}
    \caption{Ablation at matched probing budget ($K_t{=}2$). Left: locked-beam Top-1/Top-3 accuracy. Right: locked-beam SNR statistics (mean and p05) with reliability indicators (outage at $\theta$ and observed shield activation rate).}
    \label{fig:ablations}
\end{figure}

\subsection{Modality-dropout robustness}
\begin{figure*}[t]
    \centering
    \subfloat[Top-$k$ accuracy with adaptive probing ($K_t$ from entropy).]{%
        \includegraphics[width=0.40\textwidth]{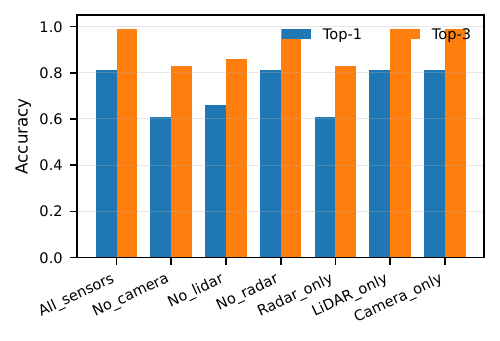}
        \label{fig:drop_topk_adapt}}
    \hfill
    \subfloat[Top-$k$ accuracy with fixed probing ($K_t\equiv2$).]{%
        \includegraphics[width=0.40\textwidth]{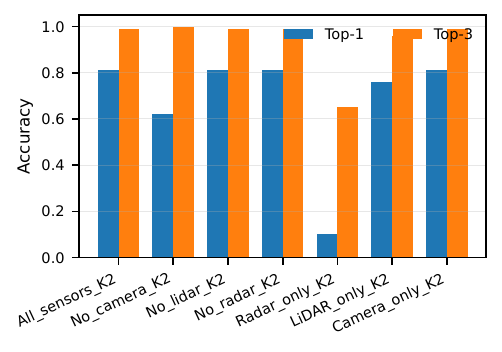}
        \label{fig:drop_topk_k2}}\\[-1mm]
    \subfloat[SNR statistics with adaptive probing (mean and p05).]{%
        \includegraphics[width=0.40\textwidth]{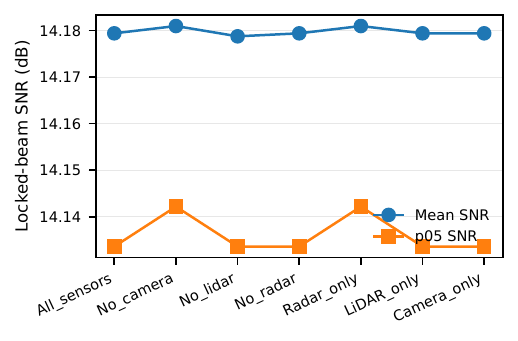}
        \label{fig:drop_snr_adapt}}
    \hfill
    \subfloat[SNR statistics with fixed probing ($K_t\equiv2$; mean and p05).]{%
        \includegraphics[width=0.40\textwidth]{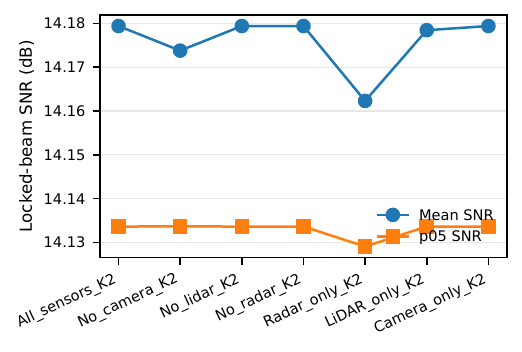}
        \label{fig:drop_snr_k2}}
    \caption{Modality-dropout robustness on DeepSense-6G scenario~43. Panels (a)–(b) report locked-beam Top-1/Top-3 accuracy across modality subsets; panels (c)–(d) report locked-beam SNR proxy statistics (mean and p05). Adaptive probing mitigates low-confidence sensing states by allocating more probes when the calibrated prior is diffuse, while the probe-and-lock step preserves high locked-beam SNR across settings.}
    \label{fig:dropout_panel}
\end{figure*}

To isolate the impact of individual sensing modalities, Table~\ref{tab:dropout_k2} reports modality-dropout results under a fixed probing budget ($K_t{=}2$), while Fig.~\ref{fig:dropout_panel} complements the table by summarizing both Top-$k$ and SNR statistics under (i) adaptive probing and (ii) fixed-$K_t$ probing.
Removing the camera substantially reduces Top-1 accuracy, and the radar-only setting degrades sharply, which indicates that a single sensing modality may be insufficient to reliably rank beams in this scenario.
At the same time, the locked-beam SNR statistics remain tightly clustered across modality subsets (Fig.~\ref{fig:dropout_panel}(c)–(d)), reflecting that the probe-and-lock step confirms the final decision using measured SNR and that many neighboring beams yield similar link quality in this test configuration.

\begin{table}[t]
\vspace{-1em} 
\centering
\caption{Modality-dropout ablation under fixed probing budget ($K_t{=}2$) on DeepSense-6G scenario~43 ($B=21$).}
\label{tab:dropout_k2}
\setlength{\tabcolsep}{6pt}
\begin{tabular}{lccccc}
\toprule
Setting & Top-1 & Top-3 & SNR$_{\rm avg}$ & SNR$_{\rm p05}$ & $K_t$ \\
\midrule
All sensors  & 0.81 & 0.99 & 14.18 & 14.13 & 2 \\
No camera    & 0.62 & 1.00 & 14.17 & 14.13 & 2 \\
No LiDAR     & 0.81 & 0.99 & 14.18 & 14.13 & 2 \\
No radar     & 0.81 & 0.99 & 14.18 & 14.13 & 2 \\
Radar only   & 0.10 & 0.65 & 14.16 & 14.13 & 2 \\
LiDAR only   & 0.76 & 0.96 & 14.18 & 14.13 & 2 \\
Camera only  & 0.81 & 0.99 & 14.18 & 14.13 & 2 \\
\bottomrule
\end{tabular}
\vspace{-1em} 
\end{table}
\subsection{Codebook-size robustness}
Table~\ref{tab:codebook_sweep} reports a uniform subcodebook sweep with $B\in\{9,13,17,21\}$.
In this dataset configuration, Top-$k$ and locked-beam SNR/outage remain unchanged across these subcodebooks, suggesting that the dominant beams are preserved under sub-sampling for this scenario.
We include this sweep to demonstrate that the pipeline executes consistently under varying $B$; broader mobility regimes and multi-user interference remain important directions for future evaluation.
\begin{table}[t]
\centering
\caption{Codebook-size sweep using uniform subcodebooks of the $B=21$ DFT codebook (adaptive policy).}
\label{tab:codebook_sweep}
\setlength{\tabcolsep}{4pt}
\begin{tabular}{lccccc}
\toprule
$B$ used & Top-1 & Top-3 & $\mathbb{E}[K_t]$ & SNR$_{\rm avg}$ & Outage \\
\midrule
9  & 0.81 & 0.99 & 1.99 & 14.18 & 0.00 \\
13 & 0.81 & 0.99 & 1.99 & 14.18 & 0.00 \\
17 & 0.81 & 0.99 & 1.99 & 14.18 & 0.00 \\
21 & 0.81 & 0.99 & 1.99 & 14.18 & 0.00 \\
\bottomrule
\end{tabular}
\vspace{-1em} 
\end{table}
\subsection{Limitations and Future Work}\label{sec:limitations}
This letter evaluates a single-link scenario on a fixed real-world dataset split.
While we strengthen breadth via LinUCB, modality dropout, and codebook sweeps, multi-user interference, broader mobility conditions, and different array configurations remain open directions.
We treat multimodal sensing measurements as a given side information and focus on resource coupling.
A natural extension is closed-loop joint sensing/communication design where sensing waveform and communication training are co-optimized, and uncertainty is propagated into both subsystems.
\section{Conclusion}
We proposed a sensing-assisted adaptive beam probing framework that combines calibrated multimodal priors with a Q-ensemble and UCB-style uncertainty-aware scheduling. 
The policy explicitly couples sensing confidence to the probing budget.
On DeepSense-6G scenario~43 with $B=21$, the proposed policy achieves Top-1/Top-3 $=0.81/0.99$ using $\mathbb{E}[K_t]=1.99$ probes per sweep, with $0\%$ outage under $\theta=0$\,dB (margin $\Delta=3$\,dB) and mean/p05 locked-beam SNR $14.18/14.13$\,dB.
Fixed-$K$ control, modality dropout, a codebook-size sweep, and a LinUCB baseline support reproducibility and practical assessment.
\section*{Acknowledgement}
This work was jointly supported by the African Center
of Excellence in Internet of Things (ACEIoT) University of
Rwanda, Regional Scholarship and Innovation Fund (RSIF),
and National Research Foundation of Korea under Grant RS-2024-00409492.
\bibliographystyle{IEEEtran}
\bibliography{IEEEabrv, References}
\end{document}